# Gridless Super-Resolution Sparse Recovery for Non-sidelooking STAP using Reweighted Atomic Norm Minimization

Tao Zhang[1], Hai Li[1], Yongsheng Hu[2], Ran Lai[1], Juncheng Guo[1]

1. Tianjin Key Laboratory for Advanced Signal Processing, Civil Aviation University of China, Tianjin, China

2. School of Information Engineering, Binzhou University, Shandong, China

**Abstract:** Sparse recovery Space-time Adaptive Processing (STAP) can reduce the requirements of clutter samples, and suppress clutter effectively using limited training samples for airborne radar. The whole angle-Doppler plane is discretized into small grid points uniformly in presently available sparse recovery STAP methods, however, the clutter ridge is not located exactly on the pre-discretized grid points in non-sidelooking STAP radar. The off-grid effect degrades the performance of STAP significantly. In this paper, a gridless sparse recovery STAP method is proposed based on reweighted atomic norm minimization, in which the clutter spectrum is precisely estimated in continuous angle-Doppler plane without resolution limit. Numerical results show that the proposed method provides an improved performance to the sparse recovery STAP methods with discretized dictionaries and STAP method utilizing atomic norm minimization.

**1. Introduction**

Space-time adaptive processing (STAP) is an effective method for clutter suppression in airborne phased array radar system [1][2][3]. The performance of the STAP filter is dependent on the estimation of the clutter-plus-noise covariance matrix (CCM). In the classical STAP method, twice of the system degree of freedom (DOFs) independent and identically (i.i.d.) training samples are need for an effective CCM estimation [4]. However, the required i.i.d. training samples are hardly obtained in the practical applications. The performance of STAP degrades significantly due to the inaccuracy of the estimated CCM, especially in non-stationary and heterogeneous environments [5], such as array geometric configuration, terrain variations and strong discrete scatters.

In order to increase the performance of STAP and reduce the number of training samples, several types of methods have been developed, and the classical STAP methods can be classified into two categories: Reduced-dimension methods and Reduce-rank methods. Reduced-dimension STAP algorithms utilize data-independent transformations to pre-filter the received signal, and the number of required training samples needed can be reduced to twice of the reduced dimension. e.g., Auxiliary Channel Processor (ACP) [6], Extended Factored Approach (EFA) [7], Joint Domain Localized (JDL) [8][9], $\sum \Delta - \text{STAP}$ [10] and Generalized Multiple Beams (GMB) [11]. Reduced-rank STAP approaches utilize a data-dependent transformations, and the number of the samples can be reduced to twice of the rank of clutter, such as Orthogonal Projection Processor (OPP) [2], Minimum Power Eigen canceller (MPE) [12], Cross-Spectral Metric (CSM) [13] and Multistage Winer Filter (MWF) [14][15][16].

In recent years, sparse recovery (SR) approach has gained widely attentions for its ability to provide innovative solutions to the problems of signal estimation, where the signal is sparse in some basis. The sparse recovery method has been successfully applied to improve the STAP performance, i.e., sparse recovery-based STAP (SR-STAP). By exploiting the intrinsic sparsity of the clutter in the angle-Doppler domain, SR-STAP methods achieve effective performance with limited training samples. Maria firstly developed a sparse recovery method to estimate target and clutter spectrum [17]. Sun proposed a method for clutter spectrum estimation using focal underdetermined system solution (FOCUSS) and $\ell_1$ norm minimization[18][19]. Li used a weighted least-squares-based iterative adaptive approach (IAA) to form angle-Doppler images of both clutter and targets [20]. For improving the reconstruction accuracy of sparse recovery, multiple measurements are applied for SR-STAP [21]. By utilizing the prior knowledge, Yang proposed a knowledge-aided space-time adaptive processing (KA-STAP) algorithm using sparse recovery [22]. Concerning the problem of array model mismatch, Yang presented a SR-STAP based on the alternating direction method to overcome the performance degradation caused by array gain and phase errors [23]. To overcome the problem of parameter-dependent in SR-STAP, Duan proposed a SR-STAP method based on sparse Bayesian learning (SBL) [24].

In the above mentioned SR-STAP methods (termed as on-grid SR-STAP), the whole angle-Doppler plane is discretized into small grid points uniformly, and the clutters are assumed to located exactly on the pre-discretized grid points of the angle-Doppler plane. The set of space-time vectors of all grid points are called space-time steering dictionary. However, the actual clutter components distribute on the clutter ridge continuously, some patches of the clutter components are not located on the girds, e.g, in non-sidelooking radar, which is called off-grid effect. The accuracy of sparse recovery is relied on the discredited space-time steering dictionary, and the off-gird effect degrades the performance of STAP significantly. To solve the off-grid problem, some improved sparse recovery methods for STAP are proposed (termed as off-grid SR-STAP). The SR-STAP with dictionary learning has been proposed in [25]. Orthogonal matching pursuit (OMP) with parameter-searching is also proposed to eliminate the effect of off-grid effect [26]. Prior knowledge of the clutter ridge is exploited to mitigate the off-grid effect for SR-STAP[27]. Although these methods improve the performance of STAP, the discredited dictionary is still needed, then the off-gird effect is not avoided. Recently, the continuous compressive sensing (CCS) is introduced for super-resolution sparse recovery, i.e., atomic $\ell_0$ norm [28]. For the Low-rank property of covariance matrix, the atomic $\ell_1$ norm, i.e., atomic norm, as a computationally feasible alternative of atomic $\ell_0$ norm, is proposed to reconstruct the atoms in a continuous-valued frequency by utilizing Vandermonde decomposition [29]. The atomic norm minimization (ANM) method is extended for 2D frequency estimation [30][31] and has been proposed for sparse-based STAP in sidelooking radar [32][33] (termed as gridless SR-STAP). However, the atomic norm suffers from a resolution limit due to the relaxation, and the frequency (normalized) estimations have to be sufficiently separated to 4/DOF (DOF denotes the degree of system freedom) for successful recovery. This prohibits high resolution recovery for clutter spectrum, especially in non-sidelooking radar, in which the clutter ridge is a curve in the continuous angle-Doppler plane and some clutter patches are separated less than the resolution limit 4/DOF. A non-convex relaxation of atomic $\ell_0$ norm is exploited to enhance the sparsity and break the resolution limit of recovery, which is termed as reweighted atomic norm minimization (RAM) [34]. Inspired by the super-resolution of reweighted atomic norm minimization, in this work, we extend the reweighted atomic norm minimization to angle-Doppler model and a gridless super-resolution sparse recovery STAP method using reweight atomic norm minimization with single and multiple measurement vectors is proposed. In the proposed method, since the CCM is estimated in the continuous angle-Doppler domain by reweighted atomic norm minimization, the off-gird problem caused by discretizing dictionary is avoid. Due to the super resolution property of reweighted atomic norm minimization, the proposed method achieves more accurate estimation of CCM than atomic norm minimization-based STAP, which has a resolution limit. Simulations are presented to demonstrate the clutter suppression performance of the proposed method, which has significant improvement than the presently available sparse recovery STAP methods.

Notations used in this paper are as follows. $\boldsymbol{A}^T$ and $\boldsymbol{A}^H$ are the matrix transpose and conjugate transpose of $\boldsymbol{A}$ respectively. $\otimes$ denotes the kronecker product. $\mathbb{R}$ and $\mathbb{C}$ denote the sets of real and complex numbers respectively. The Upper case and lower case boldface letters denote matrices and vectors respectively. $rank(\cdot)$ denotes the rank. $tr(\cdot)$ denotes the trace.

The rest of the paper is organized as follows. Section 2 introduces the signal model of STAP radar. Section 3 reviews the sparse recovery STAP approaches and then states the off-grid effect in non-sidelooking STAP radar. Section 4 details the proposed RAM-STAP method and its SDP implication. Simulated data are used to evaluate the performance of the proposed method in Section 5. Section 6 provides the summary and conclusion.

**2. Signal model**

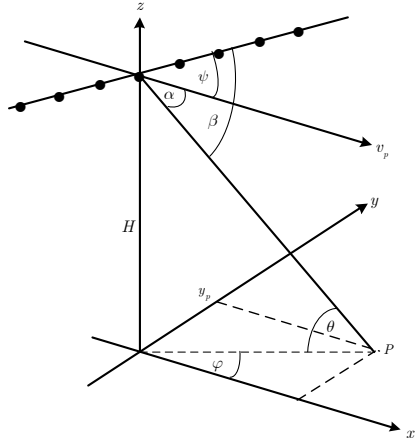

Fig.1 Airborne radar geometry with a ULA antenna

We consider a non-sidelooking uniformly linear array (ULA) airborne phased array radar, which consists $M$ antenna elements spacing half of wavelength ($d = \lambda / 2$) and $N$ pulses are received during a coherent processing interval (CPI) at a constant pulse repetition frequency (PRF) $f_r$. As shown in Fig.1, the platform is at altitude $H$ and moving with constant velocity $v_p$ along the $x$ axis. $\beta$ is the angle between clutter patch $P$ and the array line. The angles $\theta$ and $\phi$ are elevation and azimuth angles. $\psi$ denotes the crab angle between the array line and the flight direction, i.e., $\psi = 0°$ denotes the sidelooking model and $\psi = 90°$ denotes the forward-looking model.

We ignore the range ambiguities, the received target-free training sample $x \in \mathbb{C}^{NM \times 1}$ can be modeled as

$$x = x_c + n, \tag{1}$$

where $x$ is a $NM$-dimension clutter vector, and denotes as a space-time snapshot. $n$ is the thermal noise vector and $x_c$ denotes the clutter vector.

The clutter in a range ring can be model as the superposition of signals from $N_c$ independent clutter patches evenly distributed in azimuth. Then we have

$$x_c = \sum_{i=1}^{N_c} a_i s(f_{d,i}, f_{s,i}), \tag{2}$$

where $N_c$ is the number of clutter patches, $a_i$, $f_{d,i}$ and $f_{s,i}$ are the random complex amplitude, the Doppler frequency and the spatial frequency of the $i$th clutter patch. $s(f_{d,i}, f_{s,i})$ is the $NM$-dimension space-time steering vector with Doppler frequency $f_{d,i}$ and spatial frequency $f_{s,i}$, i.e., $s(f_{d,i}, f_{s,i}) = s_{d,i}(f_{d,i}) \otimes s_{s,i}(f_{s,i})$, and can be given by

$$s_{d,i}(f_{d,i}) = [1 \quad \exp(j2\pi f_{d,i}) \quad \cdots \quad \exp(j2\pi(N-1)f_{d,i})]^T, \tag{3}$$

$$s_{s,i}(f_{s,i}) = [1 \quad \exp(j2\pi f_{s,i}) \quad \cdots \quad \exp(j2\pi(M-1)f_{s,i})]^T, \tag{4}$$

where

$$f_{d,i} = \frac{2v_p}{\lambda f_r} \cos\varphi_i \cos\theta_i, \tag{5}$$

$$f_{s,i} = \frac{d}{\lambda} \cos\beta = \cos(\varphi_i - \psi). \tag{6}$$

The CCM is defined as

$$R = E[xx^H]. \tag{7}$$

Under the zero-mean complex Gaussian assumption of thermal noise vector, and based on the maximization of the signal-to-clutter-plus-noise ratio (SCNR) principle, the output of STAP with adaptive weight vector $w_{opt}$ can be given as

$$y = w_{opt}^H x, \tag{8}$$

where the adaptive weight vector calculated using the CCM

$$w_{opt}^{H} = \frac{R^{-1}s(f_{d,i}, f_{s,i})}{s^{H}(f_{d,i}, f_{s,i})R^{-1}s(f_{d,i}, f_{s,i})}. \tag{9}$$

In practice, the CCM $R$ is unknown, and can be estimated from the target-free training samples around the cell under test (CUT). Assume the clutter of training samples are independent and identically distributed with the clutter in the CUT, then the CCM $R$ can be estimated by

$$\hat{R} = \frac{1}{L}\sum_{l=1}^{L} x_l x_l^H. \tag{10}$$

Where $x_l$ denotes a target-free training sample in the $lth$ range cell. This is termed as sample matrix inversion (SMI) STAP method. However, the required number of i.i.d. training samples should be twice of the DOF to yield an average performance loss of roughly 3dB, i.e. $L > 2NM$. This is unavailable in practical environments especially for non-sidelooking radar.

### 3. SR-STAP and off-grid problem

### 3.1 Sparse recovery for STAP

Recently proposed clutter spectrum sparse recovery-based STAP methods can be summarized to two steps: clutter space-time profile is recovered by a certain sparse recovery algorithm, firstly. Then the CCM is reconstructed and the STAP weight vector is calculated from CCM, secondly.

In these methods, the angle-Doppler plane is discretised into $N_s \times N_d$ grid points, where $N_s = \rho_s N$, $N_d = \rho_d M$, $\rho_s > 1$, $\rho_d > 1$ and determine the smoothness of the angle-Doppler plane. The corresponding set of the space-time steering vectors of all grid points are formulated as

$$\Psi = [\psi_1, \psi_2, \ldots, \psi_{N_s N_d}] = S_d \otimes S_s, \tag{11}$$

where $\Psi \in \mathbb{C}^{NM \times N_s N_d}$ is termed as space-time steering dictionary, $S_d = [s_{d,1}, s_{d,2}, \ldots, s_{d,N_d}] \in \mathbb{C}^{N \times N_d}$, $S_s = [s_{s,1}, s_{s,2}, \ldots, s_{s,N_s}] \in \mathbb{C}^{M \times N_s}$, $s_{d,i}$ and $s_{s,i}$ are the space and time steering vectors defined in Eq. (3)(4).

The received target-free training samples $x$ has the form of

$$x = \Psi a + n, \tag{12}$$

where $a \in \mathbb{R}^{N_s N_d \times 1}$ is unknown solution matrix with each row representing a possible clutter patch and denotes as the clutter angle-Doppler profile, $n \in \mathbb{C}^{NM \times 1}$ denotes a Gaussian noise vector.

For the sparse property of clutter in angle-Doppler plane, the clutter angle-Doppler profile $a$ can be solved according to the theory of single measurement vector sparse recovery (SMV SR) as

$$\min_{a} \; ||a||_0 \quad \text{subject to} \quad ||\Psi a - x||_2^2 \leq \varepsilon. \tag{13}$$

where $||\cdot||_0$ denotes the $\ell_0$ norm, $||\cdot||_2$ denotes the $\ell_2$ norm and $\varepsilon$ is the noise error allowance.

After obtaining the $a$ by the sparse recovery approach, the CCM can be reconstructed by the angle-Doppler profile as

$$\boldsymbol{R} = \boldsymbol{R}_c + \boldsymbol{R}_n = \sum_{i=1}^{N_s N_d} |a_i|^2 \, \boldsymbol{\psi}_i \boldsymbol{\psi}_i^H + \sigma_n^2 \boldsymbol{I}_{NM}, \quad (14)$$

where $\sigma_n^2$ is the noise power, $\boldsymbol{I}_{NM}$ is a $NM \times NM$ identity matrix.

If multiple i.i.d. training samples are available, an improved convergence rate can be achieved by using multiple measurement vector sparse recovery (MMV SR) approach.

The received $K$ training samples $\boldsymbol{X} = [\boldsymbol{x}_1, \boldsymbol{x}_2, \ldots, \boldsymbol{x}_K] \in \mathbb{C}^{NM \times K}$ can be expressed by

$$\boldsymbol{X} = \boldsymbol{\Psi A} + \boldsymbol{N}, \quad (15)$$

where $\boldsymbol{A} = [\boldsymbol{a}_1, \boldsymbol{a}_2, \ldots, \boldsymbol{a}_K] \in \mathbb{R}^{N_s N_d \times K}$ are correlated in each nonzero row.

According to the theory of joint sparse recovery, the solution matrix $\boldsymbol{A}$ can be solved by following optimization problem:

$$\min_{\boldsymbol{A}} \quad ||\boldsymbol{A}||_{2,0} \quad \text{subject to} \quad ||\boldsymbol{\Psi A} - \boldsymbol{X}||_F^2 \leq K\varepsilon. \quad (16)$$

where $||\cdot||_{2,0}$ is the mixed norm defined as the $\ell_0$ norm of the column vector calculated by the $\ell_2$ norm of the row vectors and $||\cdot||_F$ is the Frobenius norm ($\ell_2$ norm of a matrix).

The CCM of the CUT is calculate by the recovered angle-Doppler profile as

$$\boldsymbol{R} = \boldsymbol{R}_c + \boldsymbol{R}_n = \frac{1}{K} \sum_{k=1}^{K} \sum_{i=1}^{N_s N_d} |a_{i,k}|^2 \, \boldsymbol{\psi}_i \boldsymbol{\psi}_i^H + \sigma_n^2 \boldsymbol{I}_{NM} \quad (17)$$

Since the $\ell_0$ norm minimization problem is NP-hard to solve, some relaxation are exploited and applied to SR-STAP, e.g., convex optimization $\ell_1$ norm SR-STAP[18][24] and non-convex optimization $\ell_p$ ($0 < p < 1$) norm SR-STAP[19].

**3.2 off-grid problem**

Although SR-STAP method can obtain well performance in sidelooking radar with only a few training samples (usually less than 6 samples), the clutter patches cannot always locate on the discretised grid points of the angle-Doppler plane in non-sidelooking radar, off-grid effect degrades the performance significantly [27]. The clutter ridge map of sidelooking radar in the discretised angle-Doppler plane is shown in Fig 2(a). Fig 2(b) and Fig 2(c) present the clutter ridge map of non-sidelooking radar with $\psi = 45°$ and $\psi = 90°$, respectively. In the situation of sidelooking radar, i.e., crab angle $\psi = 0°$, the clutter ridge is perfectly located on the grids of the discretised dictionary. However, in the situation of non-sidelooking radar, a lots of clutter patches locate out of the grids.

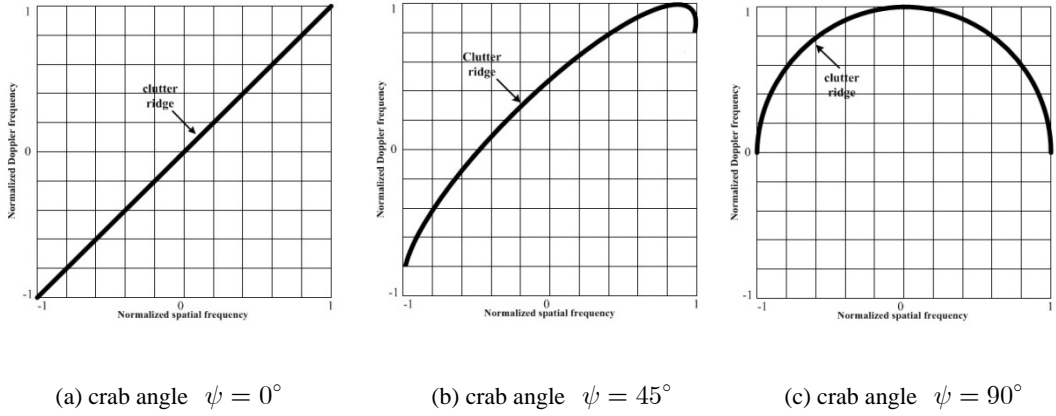

(a) crab angle $\psi = 0°$      (b) crab angle $\psi = 45°$      (c) crab angle $\psi = 90°$

Fig 2. Clutter ridge map in the discretised space-time plane

In order to tackle the off-grid problem, some dictionary learning and parameter-searching methods have been proposed [25][26], however, the discredited dictionary is still needed. A large grid interval will lead to significant error of sparse recovery, and a small grid interval results in strong column coherence of the dictionary and huge computational load. The off-gird effect is not avoided in the SR-STAP methods with discretised dictionaries.

A theory of super-resolution for frequency estimation is recently introduced by Candes, et al. [28], and a gridless convex optimization method based on atomic norm is proposed [29]. In addition, they proved that the frequencies (normalized) have to be sufficiently separated to the resolution limit (4/DOF) for successful recovery. The atomic norm minimization is extended for 2D frequency estimation [30][31] and has been exploited for sparse-based STAP in sidelooking radar[32][33] (termed as gridless SR-STAP). However, the resolution limit of atomic norm prohibits high resolution recovery for clutter spectrum and degrades the performance of clutter suppression, especially in non-sidelooking radar, in which the clutter ridge is a non-linear curve in the angle-Doppler plane and a lot of clutter patches separate less than 4/DOF.

**4. SR-STAP with reweighted atomic norm minimization**

To solve the off-grid problem and recover the clutter patches accurately, in this section, we propose a novel SR-STAP based on reweighted atomic norm minimization (RAM-STAP), which utilizes the Low-rank property of the CCM, estimates the subspace of clutter in the continuous domain without resolution limit.

For STAP radar, the clutter subspace can be spanned by $N_R$ space-time steering vectors, and the clutter covariance matrix can be decomposed as

$$\boldsymbol{R}_c = E\left[\boldsymbol{x}_c \boldsymbol{x}_c^H\right] = \sum_{i=1}^{N_R} |a_i|^2 \boldsymbol{s}(f_{d,i}, f_{s,i}) \otimes \boldsymbol{s}(f_{d,i}, f_{s,i})^H , \qquad (18)$$

where $\boldsymbol{x}_c = \sum_{i=1}^{N_R} a_i \boldsymbol{s}(f_{d,i}, f_{s,i})$ is the clutter signal, $N_R$ is the rank of the $\boldsymbol{R}_c$. According to [30], $R_c$ is a $NM \times NM$ Positive Simi-definite (PSD) Block Toeplizte matrix. The clutter patches are sparse in the angle-Doppler plane, therefore, $N_R \ll NM$, i.e., $R_c$ is a low rank matrix [35].

The set of space-time steering vectors in the continuous angle-Doppler plane can be regarded as an atomic set $\mathcal{A}$, which composes all the space-time steering vectors in the continuous angle-Doppler plane:

$$\mathcal{A} \triangleq \left\{ s(f_d, f_s) = s_d(f_d) \otimes s_s(f_s) \mid f_d \in (-1,1), f_s \in (-1,1) \right\}. \tag{19}$$

Due to the intrinsic sparsity of the clutter, the direct sparse metric of $x_c \in \mathbb{C}^{NM \times 1}$ is the atomic $\ell_0$ norm, and it is defined as

$$\left\| x_c \right\|_{\mathcal{A},0} \triangleq \inf_{f_d, f_s} \left\{ \mathcal{K} : x_c = \sum_k a_k s(f_{d,k}, f_{s,k}) \right\}. \tag{20}$$

In order to enhance the recovery resolution of clutter profile, the clutter can be recovered by the non-convex relaxation of atomic $\ell_0$ norm, i.e., reweighted atomic norm minimization as

$$\begin{aligned} &\begin{bmatrix} x_c & S(T) \end{bmatrix} = \mathcal{M}(x_c) \\ &\arg\min_u \left[ \ln \left| S(T) + \zeta I \right| + \phi \right], \\ &\text{subject to } \begin{bmatrix} \phi & x^H \\ x & S(T) \end{bmatrix} \geq 0, \left\| x_c - x \right\|_2^2 \leq \varepsilon \end{aligned}, \tag{21}$$

where $\zeta > 0$ is a regularization parameter, while $\zeta \to 0$ the reweighted atomic norm gets close to the atomic $\ell_0$ norm. $\phi \in \mathbb{C}^{1 \times 1}$ and $S(T(u))$ is a $N \times N$ block Toeplizte matrix, defined by

$$S(T(u)) = \begin{bmatrix} T_0 & T_{-1} & \cdots & T_{-(N-1)} \\ T_1 & T_0 & \cdots & T_{-(N-2)} \\ \vdots & \vdots & \vdots & \vdots \\ T_{N-1} & T_{N-2} & \cdots & T_0 \end{bmatrix}, \tag{22}$$

$T_i(u)$ is a $M \times M$ Toeplizte matrix, defined by

$$T_i(u) = \begin{bmatrix} u_{i,0} & u_{i,-1} & \cdots & u_{i,-(M-1)} \\ u_{i,1} & u_{i,0} & \cdots & u_{i,-(M-2)} \\ \vdots & \vdots & \vdots & \vdots \\ u_{i,M-1} & u_{i,M-2} & \cdots & u_{i,0} \end{bmatrix}. \tag{23}$$

The optimization function in Eq. (21) is non-convex, and can be solved by Semi-Definite Programming (SDP) with locally convergent approach Majorization-Maximization (MM) [36] algorithm.

Let $u_i$ denotes the $i$th iterate of the optimization variable $u$, then at the $(i+1)$th

iteration we have

$$\begin{bmatrix} \boldsymbol{x}_c & \boldsymbol{S}(\boldsymbol{T}(\boldsymbol{u}_{i+1})) \end{bmatrix} = \arg\min_{u} tr\left[ (\boldsymbol{S}(\boldsymbol{T}(\boldsymbol{u}_i)) + \zeta \boldsymbol{I})^{-1} \boldsymbol{S}(\boldsymbol{T}(\boldsymbol{u}_{i+1})) \right] + \phi,$$
$$\text{subject to } \begin{bmatrix} \phi & \boldsymbol{x}^H \\ \boldsymbol{x} & \boldsymbol{S}(\boldsymbol{T}(\boldsymbol{u}_{i+1})) \end{bmatrix} \geq 0, \|\boldsymbol{x}_c - \boldsymbol{x}\|_2^2 \leq \varepsilon \quad (24)$$

At each iteration, Eq. (24) can be solve by a standard SDP approach.

After enough iterations, the output $\boldsymbol{S}(\boldsymbol{T}(\boldsymbol{u}))$ denotes the subspace of clutter, then the CCM can be obtained by

$$\hat{\boldsymbol{R}}_{RAM} = \frac{1}{L} \sum_{l=1}^{L} \boldsymbol{U} diag(|\boldsymbol{U}^{-1} \boldsymbol{x}_{c,l}|^2) \boldsymbol{U}^H + \sigma_n^2 \boldsymbol{I}_{NM}. \quad (25)$$

where $\boldsymbol{S}(\boldsymbol{T}(\boldsymbol{u})) = \boldsymbol{U} \Sigma \boldsymbol{U}^{-1}$ denotes the Eigen-decomposition.

If multiple i.i.d. training samples are available, a significant improved performance can be obtained by MMV joint sparse recovery approach. We extend the reweighted atomic norm minimization to the case in presence of MMV, and a MMV-based RAM-STAP can be expressed in the following.

Let $K$ clutter signal matrix $\boldsymbol{X}_c = [\boldsymbol{x}_{c,1}, \boldsymbol{x}_{c,2}, \ldots, \boldsymbol{x}_{c,K}] \in \mathbb{C}^{NM \times K}$, and it can be decomposed as

$$\boldsymbol{X}_c = \sum_{i=1}^{N_R} \boldsymbol{s}(f_{d,i}, f_{s,i}) \boldsymbol{a}_i = \sum_{i=1}^{N_R} c_i \boldsymbol{s}(f_{d,i}, f_{s,i}) \boldsymbol{v}_i \quad (26)$$

where $\boldsymbol{a}_i = [a_1, \ldots, a_K] \in \mathbb{C}^{1 \times K}$, $c_i = \|\boldsymbol{a}_i\|_2$ and $\boldsymbol{v}_i = c_i^{-1} \boldsymbol{a}_i$ with $\|\boldsymbol{v}_i\|_2 = 1$. Let $\mathbb{Q}^{2K-1} = \{\boldsymbol{v} \in \mathbb{C}^{1 \times K} : \|\boldsymbol{v}\|_2 = 1\}$ and the set of atoms is defined as

$$\mathcal{A} \triangleq \{\boldsymbol{s}(f_d, f_s, \boldsymbol{v}) = \boldsymbol{s}(f_d, f_s) \boldsymbol{v} \mid f_d \in (-1,1), f_s \in (-1,1), \boldsymbol{v} \in \mathbb{Q}^{2K-1}\} \quad (27)$$

The atomic $\ell_0$ norm of $\boldsymbol{X}_c$ can be expressed as

$$\|\boldsymbol{X}_c\|_{\mathcal{A},0} \triangleq \inf_{f_d, f_s} \left\{ \mathcal{K} : \boldsymbol{X}_c = \sum_k c_k \boldsymbol{s}(f_{d,k}, f_{s,k}, \boldsymbol{v}), \boldsymbol{s}(f_{d,k}, f_{s,k}, \boldsymbol{v}) \in \mathcal{A}, c_k > 0 \right\} \quad (28)$$

The reweighted atomic norm minimization of $\boldsymbol{X}_c$ can be solved by following rank minimization problem:

$$\begin{bmatrix} \boldsymbol{X}_c & \boldsymbol{S}(\boldsymbol{T}) \end{bmatrix} = \mathcal{M}(\boldsymbol{X}_c)$$
$$\arg\min_{u} \left[ \ln |\boldsymbol{S}(\boldsymbol{T}) + \zeta \boldsymbol{I}| + tr(\Phi) \right],$$
$$\text{subject to } \begin{bmatrix} \Phi & \boldsymbol{X}^H \\ \boldsymbol{X} & \boldsymbol{S}(\boldsymbol{T}) \end{bmatrix} \geq 0, \|\boldsymbol{X}_c - \boldsymbol{X}\|_{2F}^2 \leq \varepsilon, \Phi = \Phi^H \quad (29)$$

where $\Phi$ is $L \times L$ Hermitian matrix, and Eq.(24) can be rewritten as

$$\begin{bmatrix} \boldsymbol{x}_c & \boldsymbol{S}(\boldsymbol{T}(\boldsymbol{u}_{i+1})) \end{bmatrix} = \arg\min_{u} tr\left[(\boldsymbol{S}(\boldsymbol{T}(\boldsymbol{u}_i)) + \zeta \boldsymbol{I})^{-1} \boldsymbol{S}(\boldsymbol{T}(\boldsymbol{u}_{i+1}))\right] + tr(\Phi),$$
$$\text{subject to } \begin{bmatrix} \Phi & \boldsymbol{x}^H \\ \boldsymbol{x} & \boldsymbol{S}(\boldsymbol{T}(\boldsymbol{u}_{i+1})) \end{bmatrix} \geq 0, \|\boldsymbol{x}_c - \boldsymbol{x}\|_2^2 \leq \varepsilon, \Phi = \Phi^H \quad (30)$$

Then, with the same operation in Eq. (25), the CCM with MMV can be estimated.

### 5. Numerical examples

In this section, we verify the advantage of the proposed RAM-STAP algorithm using simulated data. The simulated data is generated by the STAP airborne radar, which has $M = 8$ elements ULA antennas, and inter-element spacing is half wavelength ($d = \lambda/2$). The parameters in our simulations are set as follows: $PRF = 300Hz$, $N = 8$ pulses in each CPI, wavelength $\lambda = 0.667m$, platform height $H = 9000m$. The platform velocity $v_p = 50m/s$. There are $N_c = 360$ clutter patches uniformly distributed from azimuth $-\pi/2$ to $\pi/2$ on a flat ground in each range cell, the amplitudes of clutter patch follows a complex Gaussian distribution. The power of thermal noise $\delta^2 = 1$. The clutter to noise ratio (CNR) is fixed at $40dB$ as denoted in [1]. Three target-free training samples are simulated from range cell $R_0 = 20km$ and the resolution of range is $37.5m$. The proposed method is compared with the existing SR-STAP algorithms, such as FOCUSS-based STAP (FOCUSS-STAP) [19], SBL-based STAP (SBL-STAP) [24] and ANM-based STAP (ANM-STAP) [32]. In FOCUSS-STAP and SBL-STAP the resolution scales of space-time dictionary $\rho_s$ and $\rho_d$ are both set to 6. The regularization parameter $\lambda = 1 \times 10^{-4}$ in FOCUSS-STAP method. The initial regularization parameter $\lambda_0 = 1 \times 10^{-2}$ in SBL-STAP method.

We measure the performance of clutter suppression by the signal-to-clutter-plus-noise ratio (SCNR) loss, which is defined by the ratio of output SCNR to output signal-to-noise ratio (SNR).

$$SCNR_{Loss} = \frac{\sigma^2 \left|\boldsymbol{w}^H \boldsymbol{s}\right|^2}{\boldsymbol{s}^H \boldsymbol{s} \boldsymbol{w}^H \boldsymbol{R} \boldsymbol{w}}. \quad (31)$$

where $\boldsymbol{R}$ is the exact CCM and adaptive weight vector is calculated by

$$\boldsymbol{w} = \mu \hat{\boldsymbol{R}}^{-1} \boldsymbol{s}, \quad (32)$$

$\mu$ is a nonzero constant, $\hat{\boldsymbol{R}}$ is the estimated CCM and $\boldsymbol{s}$ is the space-time steering vector of the detection range cell.

### 5.1 Comparison of clutter spectrum recovery without off-grid effect

In order to demonstrate the off-grid effect and the improvement of the proposed method in off-grid situation, we compare the performance of sidelooking model (no off-grid effect) clutter spectrums estimated by FOCUSS-STAP, SBL-STAP, ANM-STAP and proposed RAM-STAP.

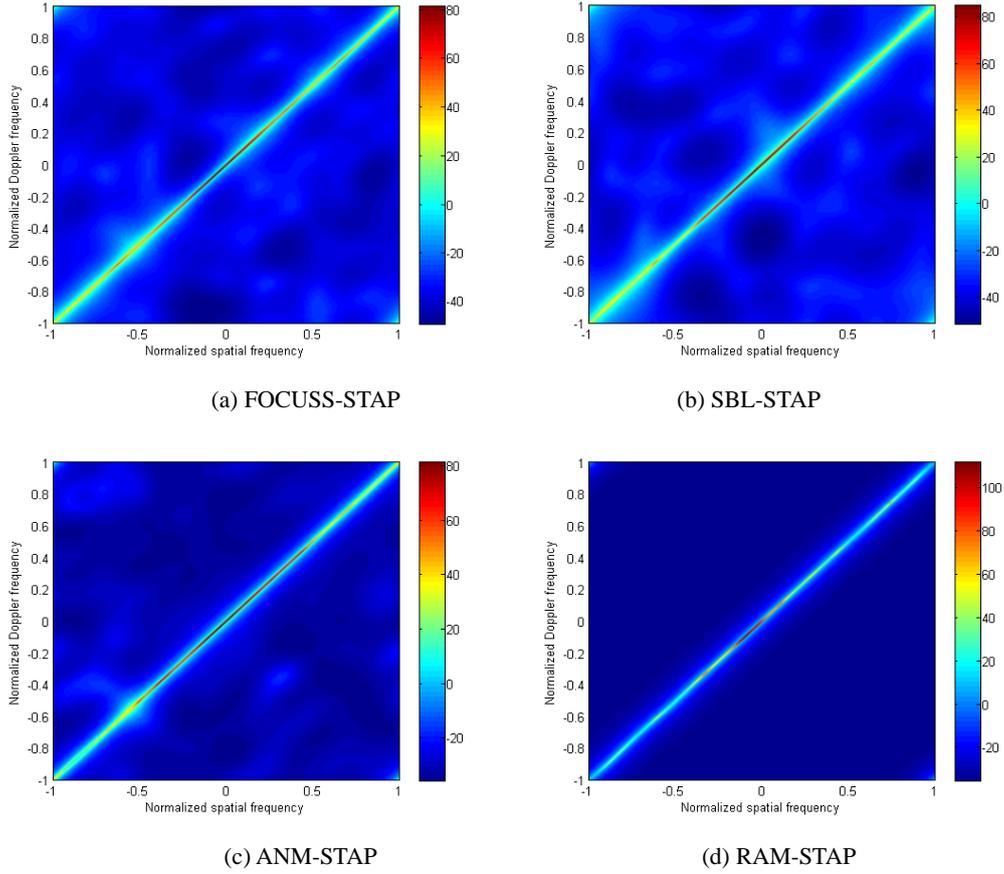

(a) FOCUSS-STAP  (b) SBL-STAP

(c) ANM-STAP  (d) RAM-STAP

Fig. 3 Recovered clutter spectrum in sidelooking array

In sidelooking array situation, the clutter patches locate on the grids points of the angle-Doppler plane, and there is no off-grid effect. Fig.3 shows the results of the recovered clutter power spectrum. Although the RAM-STAP obtain the most refined clutter spectrum, the continuous clutter spectrums without spreading are estimated by all of the above four methods. This fact reveals that the off-grid effect is the important issue for the recovery accuracy of clutter spectrum.

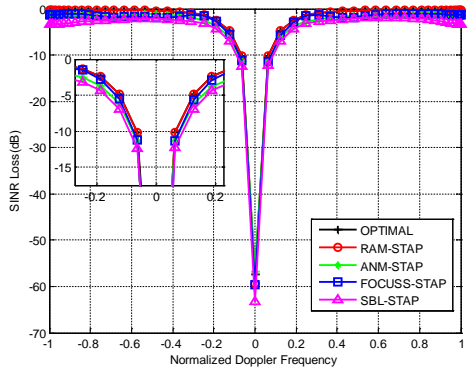 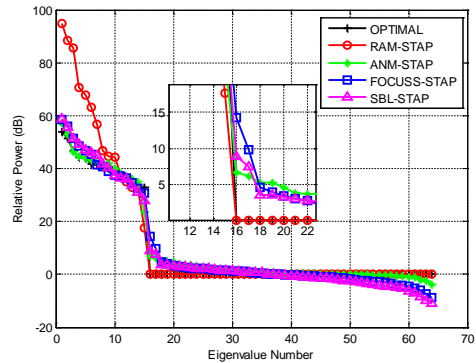

Fig.4. SINR Loss in sidelooking radar    Fig.5 Eigenspectrum of estimated CCM

in sidelooking radar

Fig.4 depicts the SINR Loss against the target Doppler frequency of the proposed method and FOCUSS-STAP, SBL-STAP, ANM-STAP methods. It is observed clearly that the SCNR loss of the proposed method and conventional SR-STAP methods are all less than 5dB except in mainlobe clutter region. Fig.5 shows the Eigenspectrum of estimated CCM, it can be seen that there are sharp cutoffs at the index 15 in the curves of all four SR-STAP methods. The results depicts that four SR-STAP methods can obtain desirable clutter spectrums while there is no off-grid effect, and the performance of SR-STAP method is mostly related to the off-grid effect.

**5.2 Comparison of clutter spectrum recovery with off-grid effect**

In this experiment, the impact of off-grid effect in non-sidelooking array is investigated. We consider two cases: (i) $\psi = 45°$. (ii) $\psi = 90°$. In the non-sidelooking array, the range dependency of the clutter should be considered and only limited clutter training samples can be used for CCM estimation. Fig.6 provides the clutter spectrums calculated by the CCM matrix in the case of $\psi = 45°$. In Fig.6(a) and (b), the clutter spectrums are calculated by the CCM estimated by the FOCUSS-STAP with 1 and 3 training samples, respectively. Obviously, FOCUSS-STAP method cannot obtain desirable clutter spectrum. This is because the dictionary in FOCUSS-STAP is discrised into grid points. In the non-sidelooking configuration the clutter patches are not located on the girds and the method is sensitive to the case in the presence of off-grid effect. The clutter spectrums in Fig.6(c) and (d) are recovered by the SBL-STAP. As expected, the SBL-STAP method also cannot obtain desirable clutter spectrum with off-grid effect. The clutter spectrums of ANM-STAP are depicted in Fig.6(e) and (f). Since the problem of off-grid is mostly related to the discrised dictionary, the continuous sparse recovery method atomic norm minimization is utilized for STAP. The clutter spectrums estimated by ANM-STAP are continuous, and much better than that of FOCUSS-STAP and SBL-STAP, but the clutter ridge is significant spreading in normalized Doppler frequency between 0.4 and 1.0, where the clutter patches separated not enough in Doppler frequency domain. The reason is that some clutter patches separated less than 4/DOF and the atomic norm suffers from the resolution limit for the convex relaxation of atomic $\ell_0$ norm. In Fig.6(g) and (h), it is observed clearly that the proposed method RAM-STAP derives accurate estimation of the clutter spectrum while existing off-grid effect. In the normalized Doppler frequency between 0.4 and 1.0, in which the clutter spectrum estimated by ANM-STAP presents a significant spreading, on the contrary, the clutter spectrum estimated by RAM-STAP can provide a more refined clutter spectrum and a high resolution and continuous clutter ridge is recovered. This can be explained by the fact that the RAM-STAP method can recover more exact clutter patches without resolution limit.

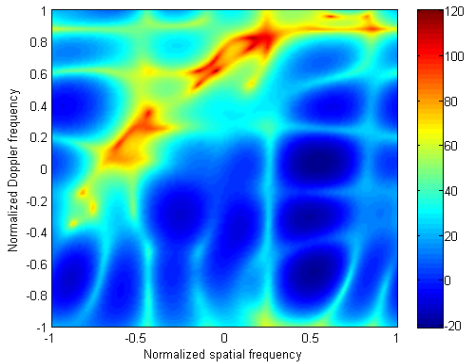
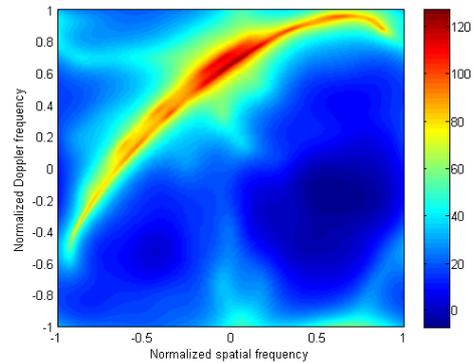

(a) FOCUSS-STAP with 1 snapshot    (b) FOCUSS-STAP with 3 snapshots

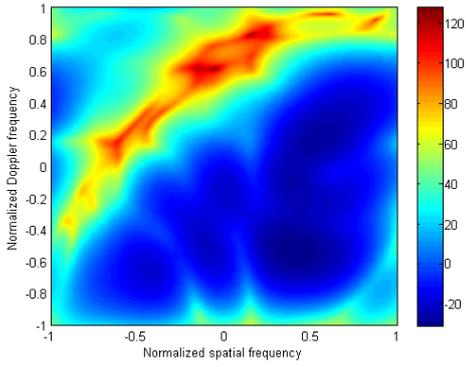

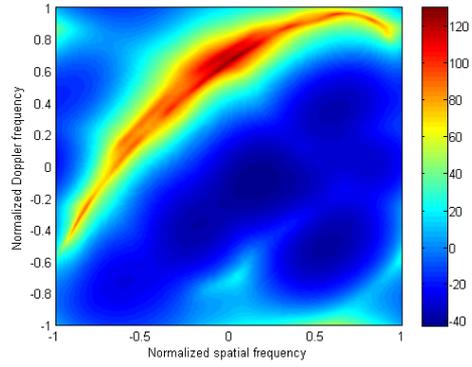

(c) SBL-STAP with 1 snapshot  (d) SBL-STAP with 3 snapshots

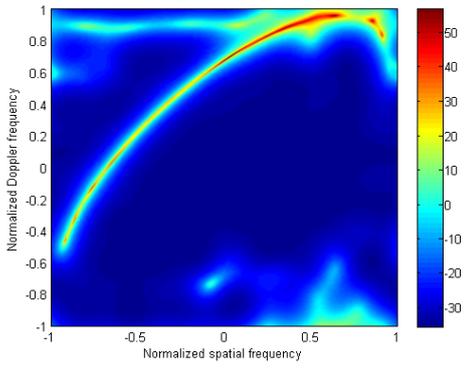

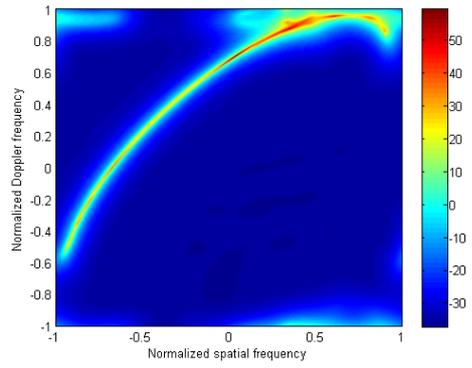

(e) ANM-STAP with 1 snapshot  (f) ANM-STAP with 3 snapshots

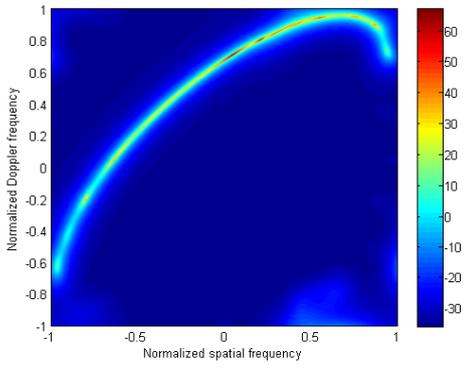

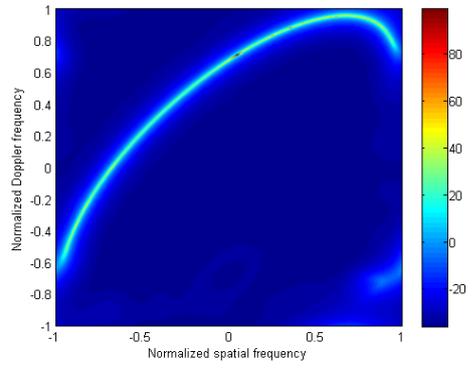

(g) RAM-STAP with 1 snapshot  (h) RAM-STAP with 3 snapshots

Fig.6 Recovered clutter spectrums in the case of $\psi = 45°$

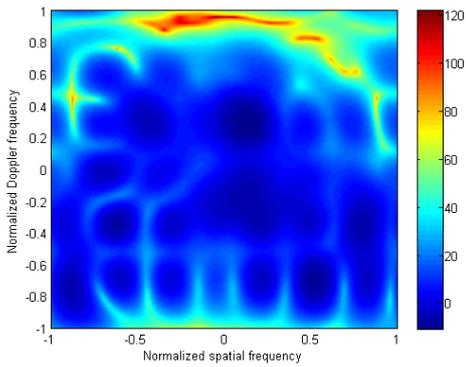

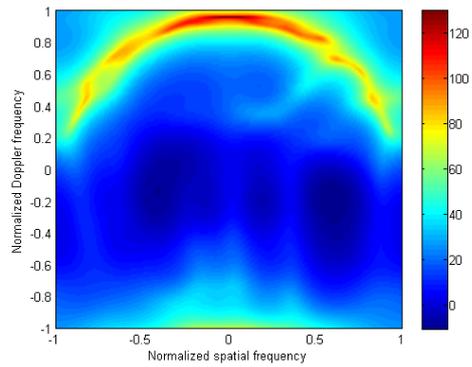

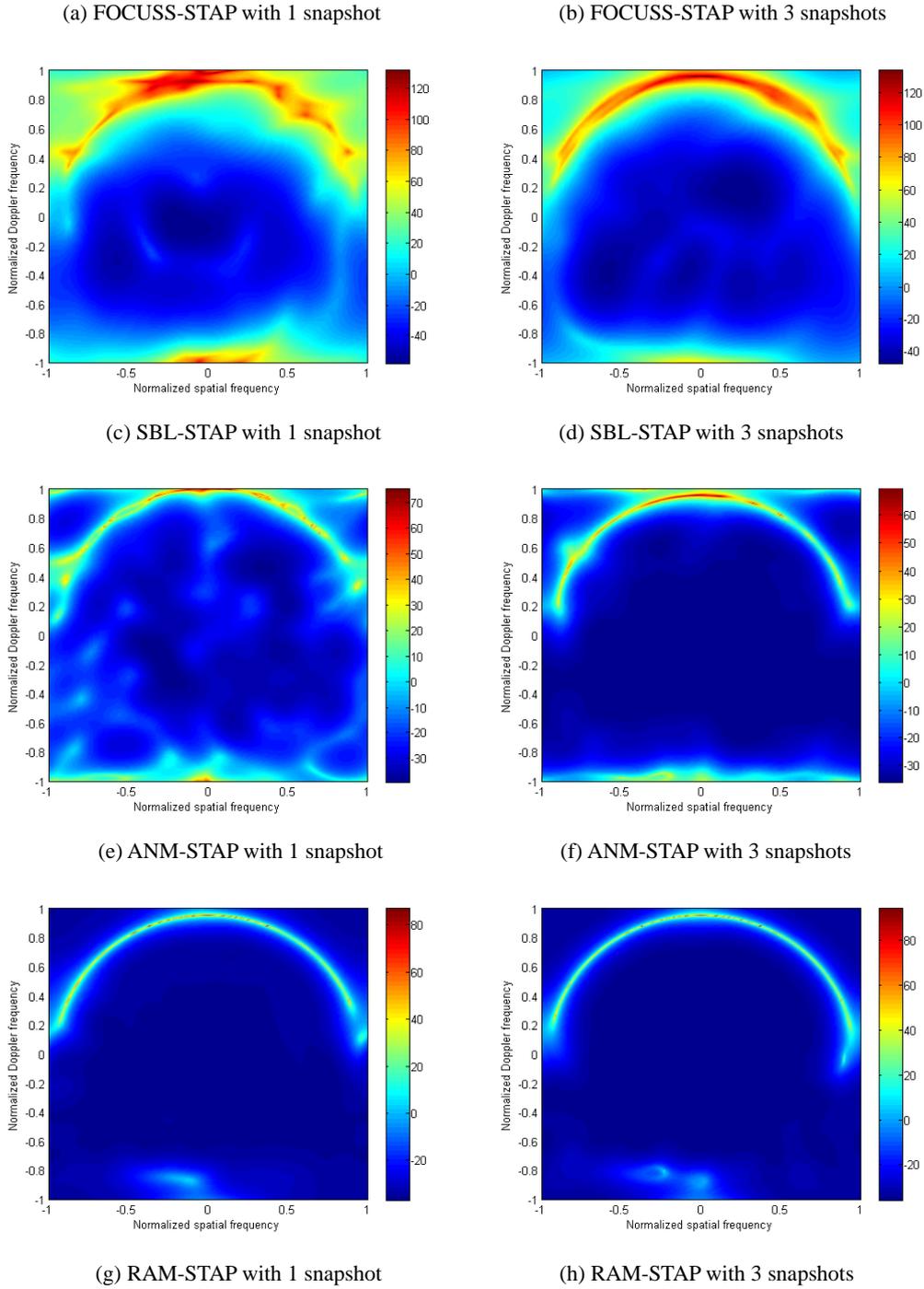

(a) FOCUSS-STAP with 1 snapshot  (b) FOCUSS-STAP with 3 snapshots

(c) SBL-STAP with 1 snapshot  (d) SBL-STAP with 3 snapshots

(e) ANM-STAP with 1 snapshot  (f) ANM-STAP with 3 snapshots

(g) RAM-STAP with 1 snapshot  (h) RAM-STAP with 3 snapshots

Fig.7 Recovered clutter spectrums in the case of $\psi = 90°$

Fig.7 presents the recovered clutter spectrum in the case of $\psi = 90°$, i.e., forwardlooking radar. In this situation, the clutter ridge is a half circle, and more clutter patches are not located on the grids. The clutter spectrums obtained via FOCUSS-STAP and SBL-STAP are depicted in Fig.7(a)-(d), where severely degradation occurs because of the off-grid effect. Although the ANM-STAP can obtain batter clutter spectrum than FOCUSS-STAP and SBL-STAP, the performance of clutter spectrum estimated by ANM-STAP degrades significantly in normalized Doppler frequency between -0.4 and 0.4. The reason is the same as in the case of $\psi = 45°$, the atomic norm minimization is the $\ell_1$ norm relaxation of atomic $\ell_0$ norm, and it tends to express the data with fewer space-time steering vectors, some actual clutter patches in the continuous clutter ridge may be lost. On the contrary, shown in Fig.7(g) and (h), the performance of proposed method has not been affected by off-grid, and the resolution of recovered clutter spectrum is still inferior to ANM-STAP method. This verifies the correctness of the results given in the case of $\psi = 45°$.

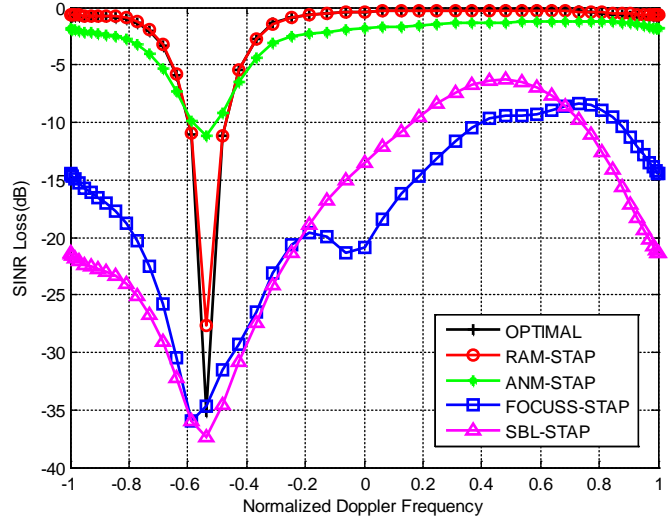

(a) in the case of $\psi = 45°$

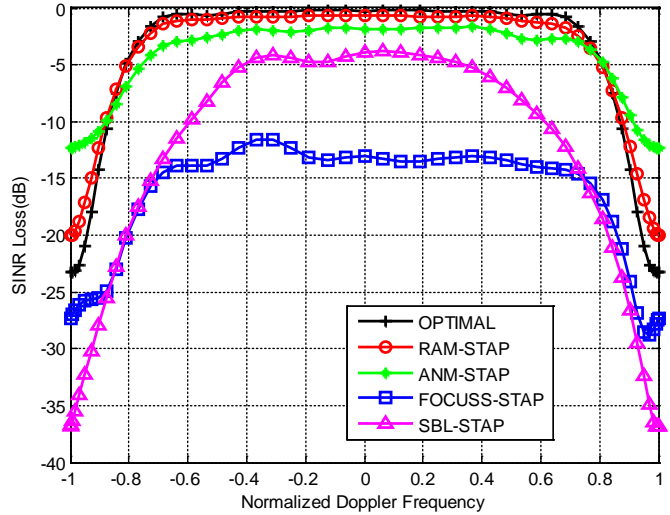

(b) in the case of $\psi = 90°$

Fig.8. SINR Loss against the target Doppler frequency in presence of off-grid

Fig.8 shows the SINR Loss of proposed method and conventional SR-STAP methods, which are averaged over 100 Monte Carlo runs. From the figures, we can see that the proposed method and ANM-STAP are robust to the cases that in presence of off-grid, but the performance of FOCUSS-STAP and SBL-STAP are significantly degraded in this case. The average SCNR loss of the proposed method is about 3dB less than the ANM-STAP and close to the OPTIMAL. In the mainlobe clutter region, RAM-STAP provides a narrower clutter notch than the ANM-STAP. When the target is away from the clutter notch, it possible to distinguish the slow-moving targets from clutter.

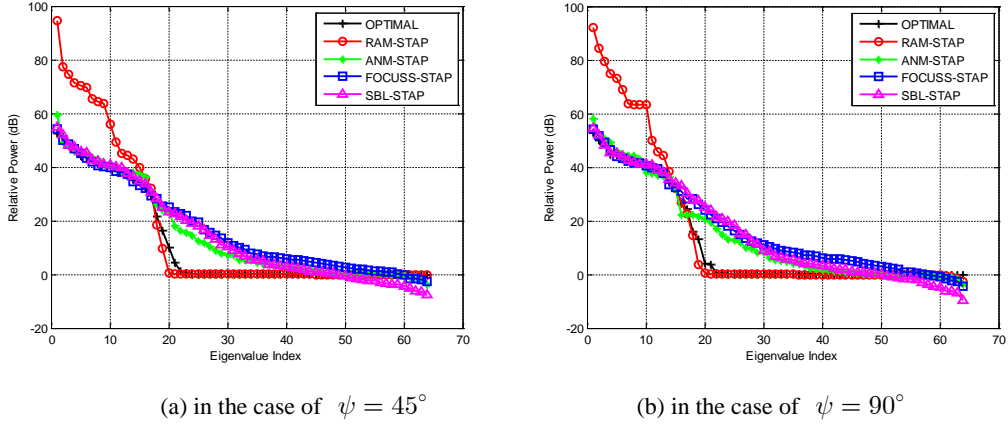

(a) in the case of $\psi = 45°$  (b) in the case of $\psi = 90°$

Fig.9. Eigenspectrum of estimated CCM

In Fig.9, we compare the eigenspectrums of the CCM estimated by different methods. Fig.9(a) and Fig.9(b) show the eigenspectrums in the case of $\psi = 45°$ and $\psi = 90°$, respectively, It is seen that the eigenvalues cutoff at the index of 22 in the eigenspectrum of the known CCM (OPTIMAL). It also can be seen that the eigenspectrum computed by the CCM estimated by the proposed RAM-STAP method closely matches the optimal eigenspectrum. In contrast, the eigenspectrums of the CCM estimated by ANM-STAP, FOCUSS-STAP and SBL-STAP methods fall off gradually and gives more significant eigenvalues than the OPTIMAL. The eigenspectrum proves that the proposed method can provide a very accurate estimation of the CCM in the case of non-sidelooking radar with off-grid effect.

## 6 Conclusion

A novel RAM-STAP method is proposed in this paper to suppress clutter for non-sidelooking airborne radar. The proposed method obtains accurate estimation of clutter spectrum by reweighted atomic norm minimization, in which the CCM is estimated in the continuous angle-Doppler domain without resolution limit. According to the simulation results, the proposed method outperforms ANM-STAP, FOCUSS-STAP and SBL-STAP methods.

## 7 Acknowledgements

This work was supported in part by the National Natural Science Foundation of China and Civil Aviation Administration of China (Grant No. U1733116), Fundamental Research Funds for Central Universities-CAUC(3122019048).